\documentclass{ws-ijmpe}

\begin{document}

\markboth{Authors' Names}{Instructions for  
Typing Manuscripts (Paper's Title)}

%
\catchline{}{}{}{}{}
%

\title{High-spin structures as the probes of proton-neutron pairing}
%

\author{A.\ V.\ Afanasjev}


\address{Department of Physics and Astronomy, Mississippi State
University, Mississippi State, P.O. Box 5167, MS 39762, USA}

%
%


\maketitle

\begin{history}
\received{(received date)}
\revised{(revised date)}
\end{history}

\begin{abstract}
Rotating $N=Z$ nuclei in the mass $A=58-80$ region have been studied within the 
framework of isovector mean field theory. Available data is well and systematically 
described in the calculations. The present study supports the presence of strong 
isovector $np$ pair field at low spin, which is, however, destroyed at high spin.
No clear evidence for the existence of the isoscalar $t=0$ $np$ pairing has been
found.
\end{abstract}

\section{Introduction}

  It is well known that in the nuclei away from the $N=Z$ line  proton-proton 
($pp$) and neutron-neutron ($nn$) pairing dominate. In the $N\approx Z$ nuclei, 
protons and neutrons occupy the same levels. Strong $np$ pair correlations are 
expected because of large spatial overlap of their wave functions. These correlations 
can be isoscalar and isovector. Figuring out their character and whether 
they form a static pair condensate (an average field) in respective channel 
has been a challenge since medium mass  $N=Z$ nuclei have come into reach of 
experiment.

   At present, the situation with the isovector $np$-pairing is most clarified. The 
strength of the isovector $np$-pairing is well defined by the isospin symmetry 
\cite{FS.99-NP}. A number of experimental observables such as binding energies of the 
$T=0$ and $T=1$ states in even-even and odd-odd $N=Z$ nuclei \cite{MFC.00,V.00,Rb74}, 
the observation of only one even-spin $T=0$ band in $^{74}$Rb \cite{Rb74} instead of 
two nearly degenerate bands expected in the case of no $t=1$ $np$-pairing 
\footnote{The lower-case letter $t$ 
is used for the isospin of the pair-field in order to avoid the confusion with the total 
isospin of the states denoted by $T$.}  {\it clearly point on the 
existence of pairing condensate in this channel}. The  analysis of pairing 
vibrations around $^{56}$Ni indicates  a  collective behavior of the isovector 
pairing vibrations but does not support any appreciable collectivity in 
the isoscalar channel \cite{bes-rev,MFCC.00}.

  On the other hand, it is still {\it an issue of debate whether the isoscalar
$np-$pair correlations lead to a pairing condensate.} The calculations 
with the realistic forces (Paris force, Argonne V14 force) indicate that the 
isoscalar pairing gap in the symmetric nuclear matter is 3 times larger than the 
isovector one \cite{GSMLSS.01}. The potential problem stems from the transition 
from realistic to effective interaction: the extremely strong $t=0$ pairing emerges 
essentially from the fact that with respect to the $t=1$ channel, dominated by 
the central force, the tensor force is acting additionally. However, the medium 
modification (screening) of the tensor force is still controversial subject 
\cite{ZZ.91}. Thus, the addition of tensor component into isoscalar pairing 
channel of the models based on effective forces may be necessary for a correct 
description of $np$-pairing in this channel. In the existing mean-field models, 
this component is neglected. An additional challenge lies in the fact that the 
strength of the effective isoscalar $t=0$ $np$-pairing is not known, and thus 
has to be defined by the comparison with experimental data. The quantity most 
frequently used for that is Wigner energy \cite{SatW.97}, but it does not provide 
a unique and reliable way to define this strength (see discussion in Ref.\ 
\cite{AF.05}).

   In a given situation, two major questions arise, namely, {\bf (i)} {\it what 
are the physical observables which are sensitive to isoscalar $np$-pair condensate} 
and {\bf  (ii)} {\it which theoretical framework has to be used for the description 
of such systems?}

 Isoscalar $np-$pairing may play a role in single-beta decay \cite{MBK.89,EPSVD.97}
(see, however, Refs.\ \cite{MPK.03,NMVPR.05}), double-beta decay \cite{PSVF.96,CHHR.99}, 
transfer reactions \cite{F.70,F.71} (see, however, Ref.\ \cite{GSW.04}), alpha decay 
and alpha correlations \cite{RSSN.98,HK.00}. However, since no symmetry-unrestricted 
mean-field (and beyond mean field) calculations of $np-$pairing, based on realistic 
effective interaction and the isospin-conserving formalism, have been carried out so 
far, no hard evidence for the elusive $t=0$ $np-$pairing phase has yet been found.
In addition, it was suggested that rotational properties (moments of inertia, band
crossing frequencies etc) can provide a signal for the
existence of  isoscalar $np$-pair condensate \cite{SatW.97,KZ.98,SW.00,SatW.00,G.01}.
These properties will be in the focus of the present manuscript. 

  In general case,  the isovector and isoscalar $np-$pairing as well as isospin symmetry 
conservation have to be taken into account in the $N\approx Z$ nuclei (see Ref.\ \cite{G.99} 
and references quoted therein). On the mean field level, the symmetry breaking in the case 
of $np-$pairing (especially of its isoscalar component because of the uncertainty with its 
strength) and isospin \cite{DH.95} can be small. In such situation, the exact methods of 
symmetry restoration by projection techniques have to be employed. Unfortunately, none of 
available theoretical tools take into account these correlations and requirements 
simultaneously reflecting the fact that such theories are extremely complicated. In 
particular, the isospin symmetry restoration in the presence of the $np-$pairing has been 
neglected in almost all theoretical studies of the $N\approx Z$ nuclei.  It is reasonable 
to expect that because of the complexity of the problem, no theoretical model, which will 
fully take into account above mentioned requirements, will be available in 
foreseeable future.

  In such situation, isovector mean field theory \cite{FS.99-NP} is a reasonable 
approximation for the study of rotational properties of the $N\approx Z$ nuclei, see 
Sect.\ \ref{Is-mean}. The present manuscript is an extension of our earlier 
systematic study of rotating $N\approx Z$ nuclei published in Ref.\ \cite{AF.05}. 
Analysis of recent experimental data in $^{72}$Kr and $^{76}$Sr within the
isovector mean field theory, combined systematic results on the $N\approx Z$ nuclei 
and the analysis of the expected situation in $^{64}$Ge will be presented in 
Sects.\ \ref{Sect-Kr72}, \ref{Sect-Sr76}, \ref{Systematics} and \ref{Ge64-sect}, 
respectively. Section \ref{Summary} summarizes our main conclusions.

\section{Isovector mean field theory}
\label{Is-mean}

   The isovector mean-field theory \cite{FS.99-NP} is used for the study 
of rotating properties of the $N\approx Z$ nuclei in the present 
manuscript. This theory assumes that there is no isoscalar 
$np-$pairing, but takes into account isovector $np-$pairing and 
isospin symmetry conservation. The later feature, treated in strong 
coupling limit, is a clear advantage 
of this approach since it is ignored in other studies. An additional 
advantage is the fact that standard mean field models with only $t=1$ 
like-particle pairing can be employed. The basis modification of these 
theories lies in adding the isorotational energy term $T(T+1)/2{\cal J}_{iso}$ 
to the total energy. Since, however, all low-lying rotational bands in even-even 
$N=Z$ nuclei have isospin $T=0$, this term vanishes. On the level of accuracy of the 
standard mean-field calculations, the restoration of the isospin symmetry (which 
takes care of the t=1 $np$ pair field) changes only the energy of the $T=1$ states 
relative to the $T=0$ states \cite{FS.99-NP}. With this in mind, the rotating 
properties were studied by means of the cranked Relativistic Hartree-Bogoliubov 
\cite{A190,CRHB,VRAL.05} (CRHB) theory.

 At high spin, the impact of $t=1$ pairing is negligible and consequently it can be 
neglected. In such situation, the isospin broken at low spin by isovector pairing is 
conserved automatically \cite{G.99}. Thus, the high spin ($I\geq 15\hbar$) states are 
systematically studied by means of the cranked Nilsson-Strutinsky (CNS) 
\cite{Beng85,A110,PhysRep} and the cranked Relativistic Mean Field (CRMF) 
\cite{KR.89,KR.93,A150} approaches, which assume zero pairing. The standard set of 
Nilsson parameters \cite{Beng85} is used in the CNS calculations. The CRMF and CRHB 
calculations have been performed with the NL3 parameterization of the RMF Lagrangian 
\cite{NL3} which provides good description of nuclear properties throughout 
nuclear chart. The D1S Gogny force \cite{D1S} and approximate particle number 
projection by means of the Lipkin-Nogami (LN) method have been used in the pairing 
channel of the CRHB theory.

  In the calculations without pairing, the shorthand notation  $[p,n]$ indicating 
the number $p(n)$ of occupied $g_{9/2}$ proton (neutron) orbitals is used for 
labeling of the configurations. In the cases when the holes in the $f_{7/2}$ subshell 
play a role, an extended shorthand notation $[(p_h)p,(n_h)n]$ with $p_h(n_h)$
being the number of proton (neutron) $f_{7/2}$ holes is used. 
The $3_i$ label is used for mixed low-$j$ $N=3$ orbitals, where subscript
$i$ indicates the position of the orbital within the specific 
signature group.

 In a number of publications it has been suggested that rotational properties 
of the $N\approx Z$ nuclei can provide evidence for the presence of a $t=0$ $np$ 
pair field \cite{SatW.97,KZ.98,SW.00,SatW.00,G.01}. However, the reasoning often 
ignored the considerable $\beta$- and  $\gamma$-softness of the nuclei in the 
mass region of interest \cite{SatW.97,KZ.98,SW.00}.
%
%
The question which physical observables of rotating nuclei may present evidence for 
the existence of the $t=0$ $np$ pair field is addressed in the present manuscript.
The size of the moment of inertia, the frequencies at which the pairs of particles 
align their angular momentum (band crossing frequencies), deformation properties, and 
unexpected mixing of configurations with a different number of quasiparticles have 
been discussed in the literature as possible indicators of $np$-pairing in rotating 
$N\approx Z$ nuclei \cite{FS.99-NP,SatW.97,KZ.98,SW.00,SatW.00,G.01,73Kr,Br70}.

\section{$^{72}$Kr nucleus}
\label{Sect-Kr72}

\begin{figure}[th]
\centerline{\psfig{file=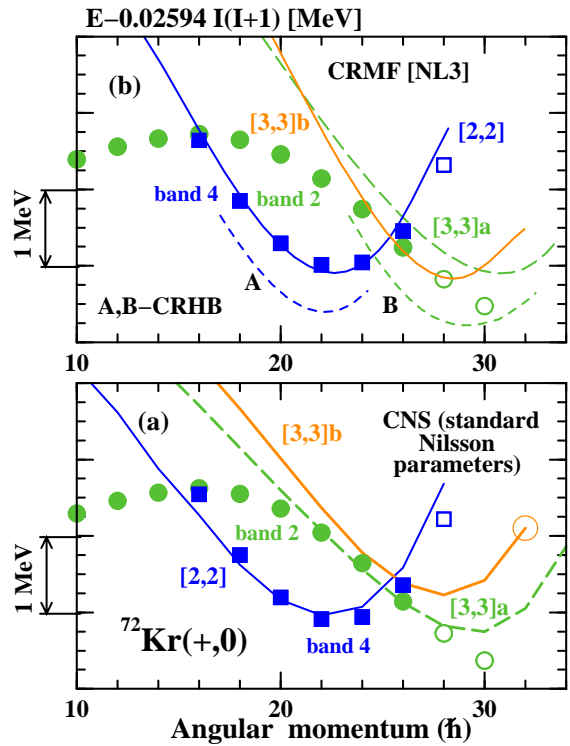,width=8cm}}
\vspace*{8pt}
\caption{Excitation energies of the experimental bands 
2 and 4 in $^{72}$Kr and theoretical configurations calculated in the CRMF, 
CRHB (panel (a)), and  CNS (panel (b)) approaches relative to a rigid rotor 
reference $E_{RLD}=E-0.02594 I(I+1)$. Experimental data are shown by symbols, 
while lines are used for theoretical results. Open symbols are used for the 
states observed recently in Refs.\ \protect\cite{Kr72a,Kr72b}. \label{E-b2-4}}
\end{figure}

  In recent experiment \cite{Kr72a,Kr72b}, previously observed bands 
were extended to an excitation energy of $\sim$24 MeV and angular momentum 
of 30$\hbar$, new side band has been observed and the lifetimes of high-spin 
states were measured for the first time. These data allow to check further
the accuracy of the description of rotating nuclei within the isovector mean
field theory. In particular, it allow to see if there is any enhancement
of the quadrupole deformation in the $N=Z$ nuclei. Ref.\ \cite{TWH.98} 
predicted that the $t=0$ $np$-pairing generates such enhancement. An important 
aspect of this study is the fact that all theoretical calculations (partially 
published in Ref.\ \cite{AF.05}; see also Ref.\ \cite{Kr72-ing} for the results 
of the CNS calculations employing different set of model parameters) were 
performed before the data became available, and, thus they 
can be considered as predictions.

\begin{figure}[th]
\centerline{\psfig{file=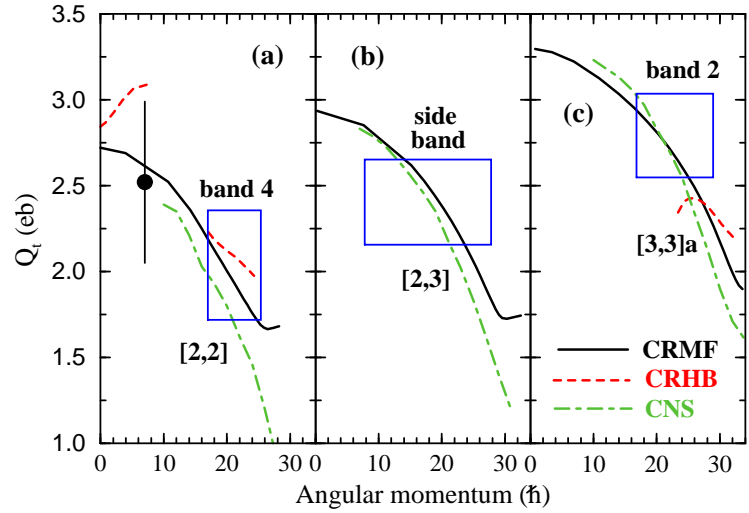,width=10cm}}
\vspace*{8pt}
\caption{Transition quadrupole moments as a function of angular momentum. The data 
point at $I = 7\hbar$ is from Ref.\ \protect\cite{Giacomo}, while boxes represent 
the measured transition quadrupole moments and their uncertainties within the 
measured spin range from the present work. The results of the CRMF and CRHB 
calculations are shown by solid and long-dashed lines, respectively.
\label{kr72-qt}}
\end{figure}

Fig.\ \ref{E-b2-4} shows the experimental excitation energies minus a rigid 
rotor reference versus angular momentum for bands 2 and 4 and the corresponding 
theoretical configurations. In Ref.\ \cite{AF.05} band 4 was assigned to the 
[2,2] configuration (i.\ e. the double S-band). This band (including recently 
observed $I=28\hbar$ state) is well described by the CNS and CRMF calculations.
These calculations also indicate the presence 
of two closely lying [3,3] configurations (Fig.\ \ref{E-b2-4}), which are the 
candidates for the band 2. The configurations [3,3]a and [3,3]b are obtained 
from the [2,2] configuration by exciting a proton and a neutron from 
the $3_3 (\alpha=-1/2)$ and $3_3 (\alpha=+1/2)$ orbitals into second 
$g_{9/2} (\alpha=+1/2)$ orbital, respectively. 
The details of the interpretation of band 2 are, however, model dependent 
reflecting the fact that the description of the energies of the single-particle 
states is not optimal (see Refs.\ \cite{PhysRep,A250}). The CNS calculations 
with the Nilsson parameters from Ref.\ \cite{GBI.86} ('A80' parameters) and the 
CRMF calculations are similar and they suggest that the band 2 may be the envelope 
of the [3,3]a and [3,3]b configurations (see top panel in Fig.\ \ref{E-b2-4}), 
whereas the CNS calculations with the standard Nilsson parameters suggest the 
[3,3]a configuration. In the former case the irregularities seen in $J^{(2)}$ of 
the band B at $\omega \geq 0.8$ MeV (see Fig.\ 2 in Ref.\ \cite{Fisher}) may be 
explained as due to the crossing (or interaction) of the [3,3]a and [3,3]b 
configurations.                 

\begin{figure}[th]
\centerline{\psfig{file=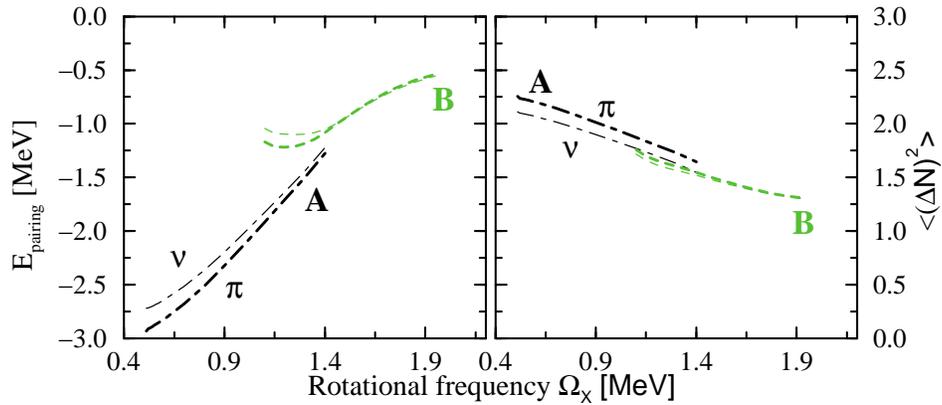,width=12cm}}
\vspace*{8pt}
\caption{Calculated values of the pairing energies  $E_{pairing}=-1/2 Tr(\Delta \kappa)$ 
and particle number fluctuations $<(\Delta N)^2>$  as a function of rotational frequency 
$\Omega_X$  in the 
CRHB configurations  A and B shown in Fig.\ \ref{E-b2-4}. The notation of
the lines is given in the figure.
\label{Pairing}}
\end{figure}

   However, a number of factors favor the assignment of the [3,3]a configuration
to the band 2. An analysis of the relative energies of experimental high-spin bands in 
$^{73,74}$Kr and $^{70}$Br shows that they are better described in the CNS 
calculations with the standard set of the Nilsson parameters as compared with the 
ones employing 
'A80' parameters. The experimental $E-E_{RLD}$ plot at spin larger than $20\hbar$ is 
better described by the [3,3]a configuration (see Fig.\ \ref{E-b2-4}). The transition 
quadrupole moment $Q_t$ of the configuration [3,3]b is smaller than the one of the 
[3,3]a configuration by 0.5-0.75 $e$b in the spin range of interest (see Fig.\ 11 in 
Ref.\ \cite{AF.05}). While the [3,3]a configuration reproduces the observed values 
of $Q_t$ of band 2 reasonably well (Fig.\ \ref{kr72-qt}), the same will not be 
possible if the configuration [3,3]b is assigned to band 2.

 The CRHB calculations were performed for the configurations A and B which 
are the paired analogs of unpaired [2,2] and [3,3]a configurations (Fig.\ \ref{E-b2-4}).
Their energies are lower than those of their unpaired analogs by approximately 0.7 MeV. 
The pairing correlations in these configurations are small (see Fig.\ \ref{Pairing}) 
and comparable with the ones in the SD band of $^{60}$Zn above the paired band  crossing 
\cite{Pingst-A30-60}. They decrease with increasing  rotational frequency reflecting 
the Coriolis anti-pairing effect. As a consequence of weak pairing correlations, the 
results of the CRHB calculations are very close to the ones of CRMF for the physical 
observables of interest such as $(E-E_{RLD})$ plots (Fig.\ \ref{E-b2-4}) (and, as a 
result, kinematic and dynamic moments of inertia), and transition quadrupole 
moments (Fig.\ \ref{kr72-qt}). It is, however, necessary to recognize that due to the 
deficiences of the Lipkin-Nogami method in the regime of weak pairing \cite{SRR.02} 
the CRMF calculations without pairing can be better approximation to exact solution 
at medium and high spins than those within the CRHB+LN framework \cite{VRAL.05}.

\begin{figure}[th]
\centerline{\psfig{file=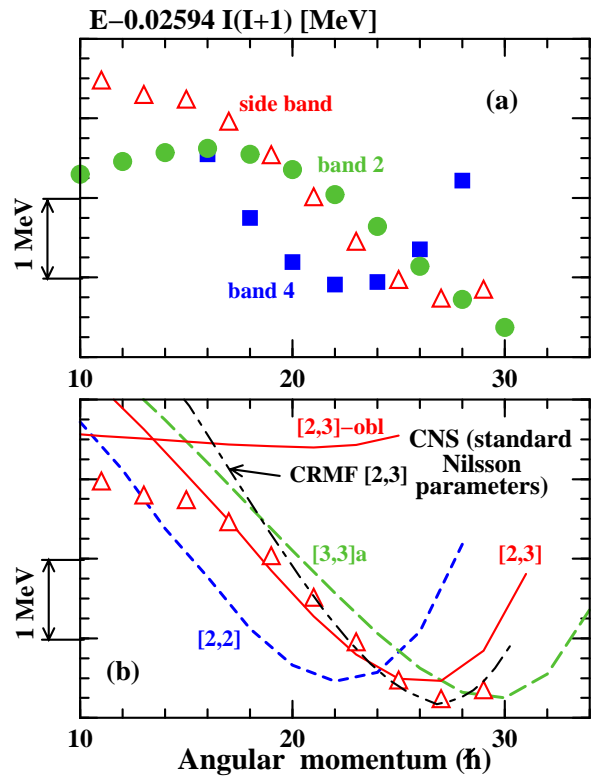,width=8cm}}
\vspace*{8pt}
\caption{ The same as Fig.\ \ref{E-b2-4}, but for the side band. Panel 
(a) shows experimental bands, while the results of the CNS calculations 
for their theoretical counterparts are given in panel (b). In addition,
the CRMF [2,3] configuration is shown in panel (b) by dash-dotted line. 
\label{Side-band}}
\end{figure}

Side band has been observed in Ref.\ \cite{Kr72a,Kr72b}. It is linked to the ground
state band by the 1685 and 1653 keV transitions of unknown multipolarity. If one assumes
E2 multipolarity for these transitions, then this band would have parity $\pi=+$ and
signature $r=0$. With this assignment, all observed high-spin bands would have the 
same parity-signature contrary to theoretical results obtained in the CNS and CRMF 
calculations which suggest the presence of near-yrast rotational sequences of negative 
parity (see Fig.\ \ref{Side-band}). Thus, E1 multipolarity is more likely choice
for the multipolarities of these transitions. With this assignment, side band has
negative parity and extends from spin $7^-$ up to spin $29^-$.  As follows from the 
$E-E_{RLD}$ plot of this band, it is built from 3 configurations with configuration 
changes (band crossings) taking place at $I=11\hbar$ and $I=17\hbar$. The high spin 
branch, which starts at $I=17\hbar$, is well described by  the [2,3] configuration 
in the CNS and CRMF calculations (Fig.\ \ref{Side-band}). At high spin, the relative 
energies of the band 4, side band and band 2  are well described in the CNS calculations 
with the standard Nilsson parameters by the [2,2], [2,3] and [3,3] configurations 
(see Fig.\ \ref{Side-band}). The calculations also suggest that low spin branch of 
side band may be associated with oblate '[2,3]-obl' configuration (Fig.\ \ref{Side-band}).

 Fig.\ \ref{kr72-qt} compares measured transition quadrupole moments of observed
bands with the ones of assigned configurations. Starting from the [2,2] configuration (band 4),
subsequent additions of the $g_{9/2}$ particle(s) increase the transition quadrupole 
moment. This trend is seen both in calculations and in experiment. In addition, absolute 
values of $Q_t$ are well described in the calculations. Experimental data 
on transition quadrupole moments are also available for  $^{73,74}$Kr \cite{Kr73Rb74-def,Kr74} 
and $^{74}$Rb \cite{Kr73Rb74-def}. These data (both absolute values and relative 
changes in $Q_t$) agree reasonably well with the results of the CNS, CRMF, and CRHB 
calculations (see Refs.\ \cite{AF.05,Kr72b,Kr73Rb74-def,Kr74} for details). In 
addition, available data on transition quadrupole moments of superdeformed rotational bands 
in $^{59}$Cu \cite{Cu59} and $^{60}$Zn \cite{A60} are well reproduced in similar calculations. 
Thus, one can conclude that {\it no enhancement of quadrupole deformation in the 
$N=Z$ nuclei (which is expected in the presence of the $t=0$ $np$-pairing \cite{TWH.98}) as 
compared with the one obtained  within the framework of isovector mean field theory
is required in order to reproduce experiment.}


\section{$^{76}$Sr nucleus: probing Coulomb antipairing effect.}
\label{Sect-Sr76}

 The progress in understanding of $np$-pairing requires better knowledge 
of different components of like-particle pairing. The investigation of 
the impact of the Coulomb exchange term on the pairing field, within the 
framework of the Hartree-Fock-Bogoliubov approach  based on the Gogny force, 
found a considerable decrease of the proton pairing energies due to a Coulomb 
anti-pairing effect \cite{Gogny}. Recent experimental data on $^{76}$Sr 
\cite{Sr76}, when combined with the limited results from previous high spin 
studies of other $A=58-80$ $N=Z$ nuclei in which $g_{9/2}$ proton/ neutron 
paired band crossings have been observed ($^{60}$Zn \cite{Zn60SD}, $^{68}$Se 
\cite{fisch03}, $^{72}$Kr \cite{kel02}) provide us with the first real 
opportunity to test these predictions \cite{Sr76}.

\begin{figure}[th]
\centerline{\psfig{file=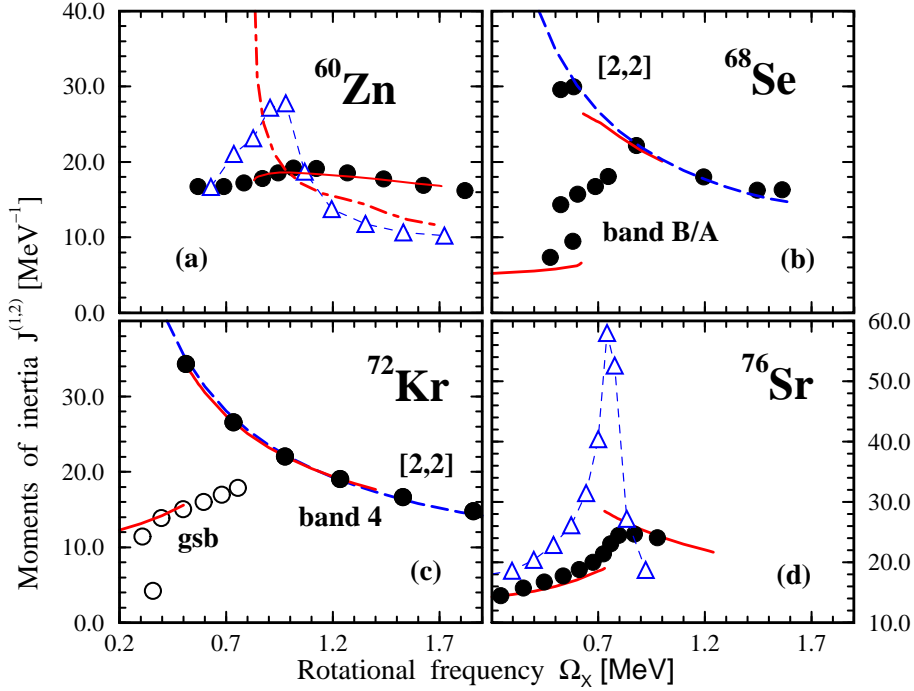,width=12cm}}
\vspace*{8pt}
\caption{Kinematic and dynamic moments of inertia of rotational bands in even-even
$N=Z$ nuclei. The dynamic moments of inertia (open triangles) are shown only for
$^{60}$Zn and $^{76}$Sr. The results of the CRHB calculations are shown by solid
and dash-dotted lines for kinematic and dynamic moments of inertia, respectively.
The kinematic moments of inertia obtained in the CRMF calculations are shown
by dashed lines in $^{68}$Se and $^{72}$Kr. 
\label{Coulomb}}
\end{figure}

%
%
 The similarity of the proton and neutron single-particle 
spectra (apart from some constant shift in absolute energies by the Coulomb energy) 
in the $N=Z$ nuclei leads to the fact that proton and neutron pairing energies are almost the same 
for proton and neutron subsystems in calculations which do not contain a Coulomb 
exchange term (as is the case with CRHB calculations, see Fig.\ \ref{Pairing}).
As a consequence, the alignment (paired band crossing) of proton and neutron pairs 
takes place at the same rotational frequency in such calculations (see Fig.\ \ref{Coulomb}), 
which in turn leads to only one bump in the dynamic moment of inertia. However, if the 
predictions of Ref.\ \cite{Gogny} are correct then the proton pairing energy should be 
considerably smaller than that due to the neutrons, and it is reasonable to expect that this
fact will result in an alignment of proton  and neutron pairs at different frequencies, 
which would manifest itself in a double peaked shape for the dynamic moments of inertia.

\begin{figure}[th]
\centerline{\psfig{file=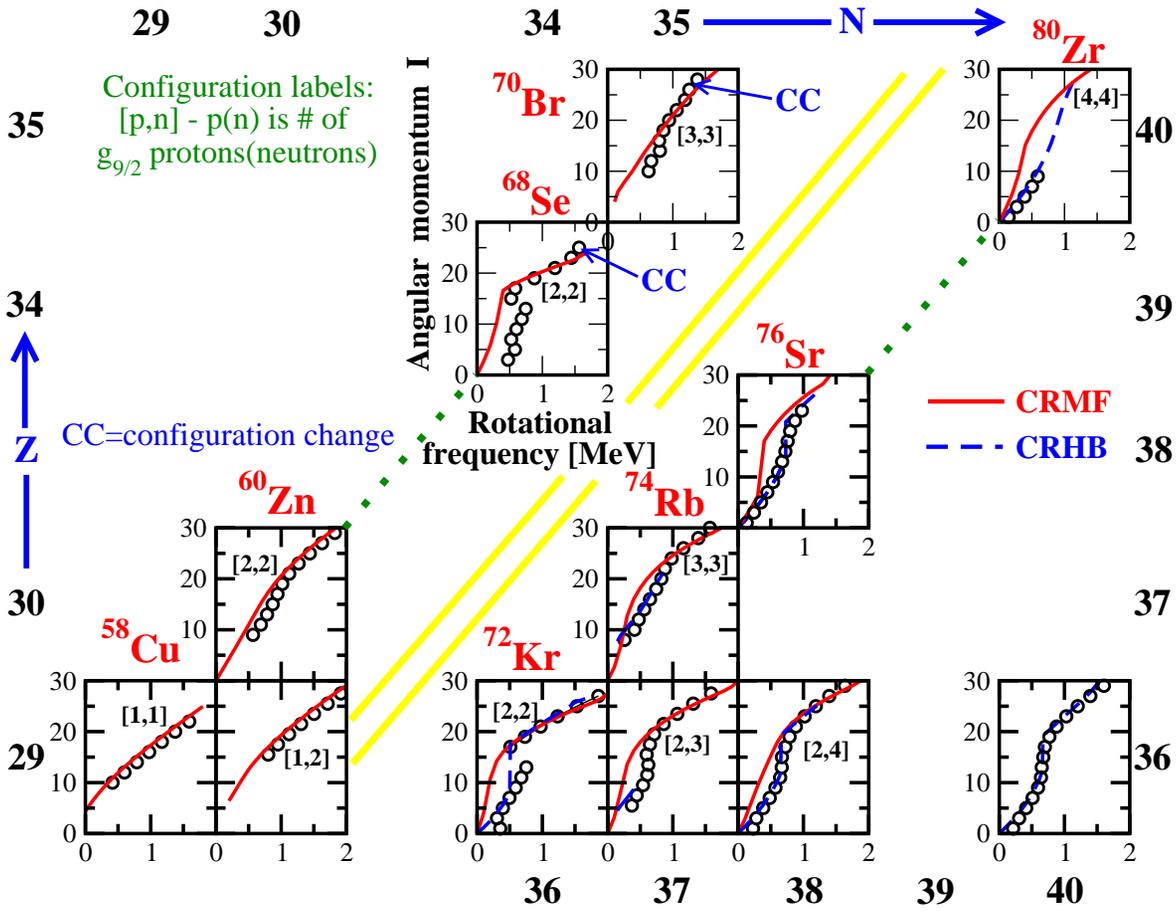,width=12.7cm}}
\vspace*{8pt}
\caption{Alignments of rotational structures in the $N\approx Z$ nuclei 
compared with the results of the CRMF and CRHB calculations. The figure 
is based on the results published in Refs.\ 
\protect\cite{Rb74,AF.05,Br70,A60,Kr73,Cu59,Sr76}. The label 'CC' indicates 
the configuration change within the configuration.
\label{syst}}
\end{figure}

 Fig.\ \ref{Coulomb} 
shows the kinematic moments of inertia for all four nuclei and the dynamic moments 
of inertia for $^{60}$Zn and $^{76}$Sr. The ground and $I^{\pi}=2^+$ states of 
$^{68}$Se and $^{72}$Kr are believed to be oblate (see Ref.\ \cite{AF.05})
which leads to low values for the kinematic moments of inertia at low frequencies 
(see Fig.\ \ref{Coulomb}). With increasing spin highly-triaxial ($^{68}$Se) or near-prolate 
($^{72}$Kr) structures become yrast \cite{AF.05}. Thus, the first irregularity seen in 
the kinematic moments of inertia of these nuclei at a rotational frequency 
$\hbar\omega \sim 0.4$ MeV is due to this shape coexistence. However such shape 
coexistence is not present in $^{60}$Zn and $^{76}$Sr at low spin. These nuclei are 
characterized by gradually increasing kinematic moments of inertia at low rotational 
frequency (see Fig.\ \ref{Coulomb}). In $^{68}$Se the band crossing seen at 
$\hbar\omega \sim 0.7$ MeV is not related to the standard change from the ground ($g-$) 
band to the $g_{9/2}$ aligned proton/ neutron $S-$band \cite{AF.05}, and, thus, 
can be excluded from consideration. It is clear, however, that for the other three 
nuclei the proton and neutron $g_{9/2}$ paired band crossings take place 
simultaneously at $\hbar\omega$ = 0.6 -- 1.0 MeV. As a result, the currently 
available experimental data in even-even $N=Z$ nuclei {\it do not support the 
existence of the Coulomb anti-pairing effect caused by the Coulomb exchange term.}
These data are also well described in the CRHB and CRMF (above paired band crossing)
calculations (see Fig.\ \ref{Coulomb} and Refs.\ \cite{AF.05,Sr76,Pingst-A30-60}
for details).

\section{Systematics of rotational and deformation properties}
\label{Systematics}

 The experimental data on alignments in rotational structures of
the $N\approx Z$ nuclei are compared with the results of the CRMF 
and CRHB calculations in Fig.\ \ref{syst}. One can see good 
agreement between experiment and the CRMF calculations at high spin 
as well as between experiment and CRHB calculations at low spin.
In addition, experimental data on transition quadrupole moments are 
available for $^{59}$Cu \cite{Cu59}, $^{60}$Zn \cite{A60}, $^{72,73,74}$Kr
\cite{Kr72a,Kr72b,Kr73Rb74-def,Kr74} and $^{74}$Rb \cite{Kr73Rb74-def}. 
These data agree well with the results of the CNS, CRMF, and 
CRHB calculations (see Sect.\ \ref{Sect-Kr72} in the present manuscript and 
Refs.\ \cite{AF.05,Kr72b,A60,Cu59,Kr73Rb74-def,Kr74} for details).

\begin{figure}[th]
\centerline{\psfig{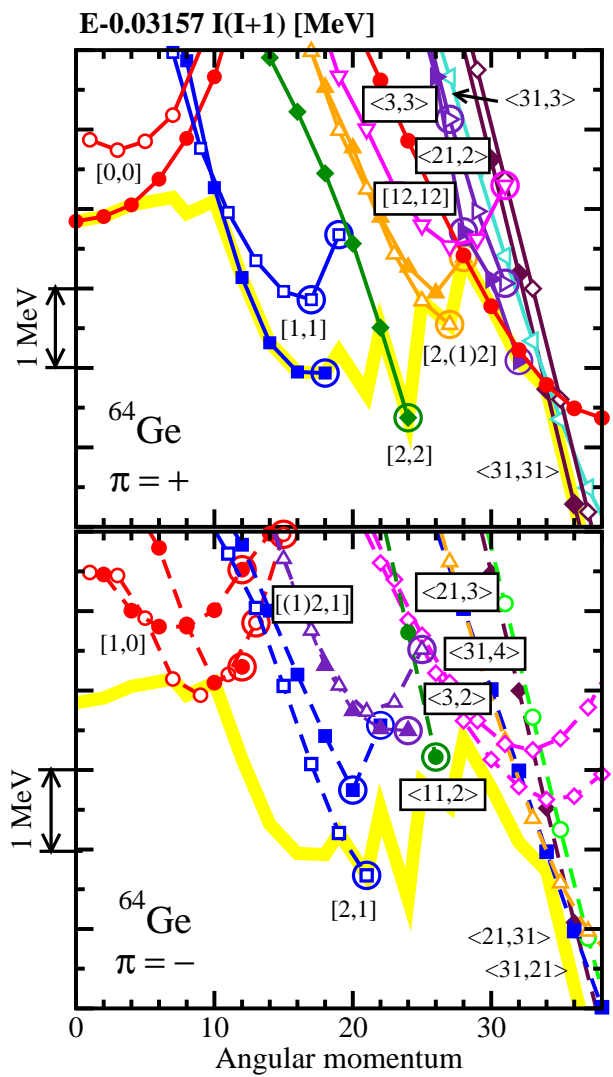}}
\vspace*{8pt}
\caption{Excitation energies of the configurations forming the yrast 
lines of 4 combinations of parity and signature in $^{64}$Ge given 
relative to a rigid rotor reference $E_{RLD}=0.03157 I(I+1)$ MeV. 
Calculated  terminating (aligned) states are encircled. The shorthand 
notation $<p_1p_2,n_1n_2>$ indicates the number $p_1(n_1)$ of occupied 
$g_{9/2}$ proton (neutron) orbitals and the number $p_2(n_2)$ of occupied 
$h_{11/2}$ proton (neutron) orbitals. $p_2(n_2)$ are omitted
when later orbitals are not occupied. Wide line indicate the total
yrast line. The same type of symbols is used for signature partner
orbitals. Solid (open) symbols are used for $\alpha=0 (1)$ 
configurations. \label{Ge64-eld}}
\end{figure}

\section{$^{64}$Ge nucleus}
\label{Ge64-sect}
 
   As follows from Fig.\ \ref{syst}, the knowledge of the $N=Z$ $^{62}$Ga, 
$^{64}$Ge, $^{66}$As, and $^{78}$Yb nuclei is restricted to low-spin states. 
The CNS calculations for $^{64}$Ge are performed in order to better 
understand what can be expected at high spin in these nuclei. 

  It has been pointed out before that $^{64}$Ge nucleus is soft with respect to
$\gamma-$ and octupole deformations (see Ref.\ \cite{Ge64-exp} and references 
quoted therein). Fig.\ \ref{Ge64-eld} shows the results of the CNS calculations
(which are restricted to reflection symmetric shapes), which also indicate  softness  
toward  $\gamma$-deformation. Indeed, the 
[0,0]$(\alpha=0)$ configuration is characterized by the $(\varepsilon_2 \sim 0.2,
\gamma \sim  -30^{\circ})$ deformation in the spin range $I=2-8\hbar$. The yrast 
lines of other combinations of parity and signature are characterized by similar 
deformation in the spin range $I=0-3\hbar$. Up to spins $I\sim 35\hbar$, the
yrast lines are dominated by the states with the deformations $\varepsilon_2 
\approx 0.25-0.35$ and $\gamma=26^{\circ}-60^{\circ}$. The terminating bands,
many of which terminate in a favored way \cite{PhysRep}, dominate the yrast
region up to $I\sim 35\hbar$. Superdeformed bands with deformation 
$\varepsilon_2 \sim 0.5, \gamma \sim 10^{\circ}$ become yrast above that spin.

  It seems that the complicated structure of this nucleus, dominated in the 
spin region of interest by $\gamma-$ (and probably octupole) softness and terminating 
structures [which are not that different from the ones in $^{68}$Se (see Fig.\ 7
in Ref.\ \cite{AF.05}) and in $^{74}$Kr (see Fig.\ 3 in \cite{AF.05})], will not 
allow to obtain reliable evidences of the isoscalar $t=0$ $np-$pairing even if 
the experimental data will be extended to higher spin.

\section{Conclusions}
\label{Summary}

  The systematic analysis of the rotational response and deformation 
properties of the $N\approx Z$ nuclei support the interpretation of 
these nuclei within the isovector mean field theory. According to this 
framework, there is no isoscalar $np$ pair field. At low spin, strong 
isovector pair field exists, which includes a large $np$ component, the 
strength of which is determined by isospin conservation. Like in nuclei 
away from the $N=Z$ line, this isovector 
pair field is destroyed by rotation. In this high spin regime the calculations 
without pairing describe well the data provided the drastic shape changes 
that cause among other things band termination are taken into account. No clear 
evidence for the existence of the isoscalar $t=0$ $np$ pairing has been found.
However, due to limitations of our theoretical tools one cannot completely 
exclude the possibility of the existence of $np$ pairing condensate
in the $t=0$ channel or the possibility that the rotational properties
are not sensitive to this type of pairing.

\section*{Acknowledgements}

  The work was supported by the DOE grant DE-FG02-07ER41459. I would like to
express my gratitude to S.\ Frauendorf, R.\ Wadsworth, and C.\ Andreoiu for
their contributions to this project.


\begin{thebibliography}{0}

\bibitem{FS.99-NP} S.\ G.\ Frauendorf and J.\ A.\ Sheikh, {\it Nucl.\ Phys.} {\bf A645}, 509 (1999).

\bibitem{MFC.00} A.\ O.\ Macchiavelli, P.\ Fallon, R.\ M.\ Clark, M.\ Cromaz, 
M.\ A.\ Deleplanque, R.\ M.\ Diamond, G.\ J.\ Lane, I.\ Y.\ Lee, F.\ S.\ Stephens, 
C.\ E.\ Svensson, K.\ Vetter, and D.\ Ward, {\it Phys.\ Rev.} {\bf C61}, 041303(R) 
(2000). 

\bibitem{V.00} P.\ Vogel, {\it Nucl.\ Phys.} {\bf A662}, 148 (2000).

\bibitem{Rb74} C.\ D.\ O'Leary, C.\ E.\ Svensson, S.\ G.\ Frauendorf,
A.\ V.\ Afanasjev, D.\ E.\ Appelbe, R.\ A.\ E.\ Austin, 
G.\ C.\ Ball, J.\ A.\ Cameron, R.\ M.\ Clark, M.\ Cromaz, 
P.\ Fallon, D.\ F.\ Hodgson, N.\ S.\ Kelsall, A.\ O.\ 
Macchiavelli, I.\ Ragnarsson, D.\ Sarantites, J.\ C.\ 
Waddington and R.\ Wadsworth, {\it Phys.\ Rev.} {\bf C67}, 021301(R) (2003).

\bibitem{bes-rev} D.\ R.\ Bes, R.\ A.\ Broglia, O.\ Hansen, O.\ Nathan, 
{\it Phys.\ Rep.} {\bf 34}, 1 (1977). 

\bibitem{MFCC.00}  A.\ O.\ Macchiavelli, P.\ Fallon, R.\ M.\ Clark, M.\ Cromaz, 
M.\ A.\ Deleplanque, R.\ M.\ Diamond, G.\ J.\ Lane, I.\ Y.\ Lee, F.\ S.\ Stephens, 
C.\ E.\ Svensson, K.\ Vetter, and D.\ Ward, {\it Phys.\ Lett.} {\bf B480}, 1 (2000).

\bibitem{GSMLSS.01}  E.\ Garrido, P.\ Sarriguren, E.\ Moya de Guerra,
U.\ Lombardo, P.\ Schuck, H.\ J.\ Schulze, {\it Phys.\ Rev.} {\bf C63}, 037304 
(2001). 

\bibitem{ZZ.91} D.\ C.\ Zheng and L.\ Zamick, {\it Ann.\ Phys.} (N.Y.) {\bf 206}, 
106 (1991).

\bibitem{SatW.97} S.\ Satu{\l}a and R.\ Wyss, {\it Phys.\ Lett.} {\bf B393}, 1 (1997).

\bibitem{AF.05} A.\ V.\ Afanasjev and S.\ Frauendorf, {\it Phys.\ Rev.} {\bf C71},
064318 (2005). 

\bibitem{MBK.89} K.\ Muto, E.\ Bender, and H.\ V.\ Klapdor, Z.\ Phys. {\bf A334},
47 (1989).

\bibitem{EPSVD.97} J.\ Engel, S.\ Pittel, M.\ Stoitsov, P.\ Vogel, and 
J.\ Dukelsky, {\it Phys.\ Rev.} {\bf C55}, 1781 (1997).

\bibitem{MPK.03} P.\ M\"oller, B.\ Pfeiffer, and K.-L.\ Kratz,
{\it Phys.\ Rev.} {\bf C67}, 055802 (2003). 

\bibitem{NMVPR.05} T.\ Nik\v{s}i\'{c}, T.\ Marketin, D.\ Vretenar, N.\ Paar, and 
P.\ Ring,  {\it Phys.\ Rev.} {\bf C71}, 014308 (2005).

\bibitem{PSVF.96} G.\ Pantis, F.\ Simkovic, J.\ D.\ Vergados,
A.\ Faessler, {\it Phys.\ Rev.} {\bf C53}, 695 (1996).

\bibitem{CHHR.99} O.\ Civitarese, P.\ O.\ Hess, J.\ G.\ Hirsch, and M.\ Reboiro,
{\it Phys.\ Rev.} {\bf C59}, 194 (1999).

\bibitem{F.70} P.\ Fr{\"o}brich, {\it Z.\ Phys.} {\bf 236}, 153 (1970).

\bibitem{F.71} P.\ Fr{\"o}brich, {\it Phys.\ Lett.} {\bf B37}, 338 (1971). 

\bibitem{GSW.04}  S.\ G{\l}owacz, W.\ Satu{\l}a, and R.\ A.\ Wyss, 
{\it Eur.\ Phys.\ J.} {\bf A19}, 33 (2004).

\bibitem{RSSN.98} G.\ R\"opke, A.\ Schnell, P.\ Schuck, and P.\ Nozi\`{e}res,
{\it Phys.\ Rev.\ Lett.} {\bf 80}, 3177 (1998).

\bibitem{HK.00} M.\ Hasegawa and K.\ Kaneko, {\it Phys.\ Rev.} {\bf C61}, 037306
(2000).

\bibitem{KZ.98} K.\ Kaneko and J.\ Zhang, {\it Phys.\ Rev.} {\bf C57}, 1732 (1998).

\bibitem{SW.00} J.\ A.\ Sheikh and R.\ Wyss, {\it Phys.\ Rev.} {\bf C62}, 051302(R) (2000).

\bibitem{SatW.00} S.\ Satu{\l}a and R.\ Wyss, {\it Nucl.\ Phys.} {\bf A676}, 120 (2000).

\bibitem{G.01} A.\ L.\ Goodman, {\it Phys.\ Rev.} {\bf C63}, 044325 (1999).

\bibitem{73Kr} N.\ S.\ Kelsall, S.\ M.\ Fischer, D.\ P.\ Balamuth, G.\ C.\ Ball, 
M.\ P.\ Carpenter, R.\ M.\ Clark, J.\ Durell, P.\ Fallon,  S.\ J.\ Freeman, 
P.\ A.\ Hausladen, R.\ V.\ F.\ Janssens, D.\ G.\ Jenkins, M.\ J.\ Leddy, 
C.\ J.\ Lister, A.\ O.\ Macchiavelli, D.\ G.\ Sarantites, D.\ C.\ Schmidt, 
D.\ Seweryniak, C.\ E.\ Svensson, B.\ J.\ Varley, S.\ Vincent, R.\ Wadsworth,
A.\ N.\ Wilson, A.\ V.\ Afanasjev, S.\ Frauendorf, I.\ Ragnarsson and 
R.\ Wyss, {\it Phys.\ Rev.} {\bf C65}, 044331 (2002).

\bibitem{G.99}  A.\ L.\ Goodman, {\it Phys.\ Rev.} {\bf C60}, 014311 (1999). 

\bibitem{DH.95}  J.\ Dobaczewski and I.\ Hamamoto, {\it Phys.\ Lett.} {\bf B345}, 181 (1995).

\bibitem{Kr72a} C.\ Andreoiu, C.\ E.\ Svensson, R.\ A.\ E.\ Austin, M.\ P.\ Carpenter,
D.\ Dashdorj, P.\ Finlay, S.\ J.\ Freeman, P.\ E.\ Garrett, A.\ G{\"o}rgen, J.\ Greene,
G.\ F.\ Grinyer, B.\ Hyland, D.\ Jenkins, F.\ Johnston-Theasby, P.\ Joshi, 
A.\ O.\ Macchiavelli, F.\ Moore, G.\ Mukherjee, A.\ A.\ Phillips, W.\ Reviol, D.\ G.\ 
Sarantites, M.\ A.\ Schumaker, D.\ Seweryniak, M.\ B.\ Smith, J.\ J.\ Valiente-Dobon,
and R.\ Wadsworth, {\it Phys.\ Scripta} {\bf T125}, 127 (2006).

\bibitem{Kr72b} C.\ Andreoiu, C.\ E.\ Svensson, A.\ V.\ Afanasjev, 
R.\ A.\ E.\ Austin, M.\ P.\ Carpenter, 
D.\ Dashdorj, P.\ Finlay, S.\ J.\ Freeman, P.\ E.\ Garrett, J.\ Greene, G.\ F.\ Grinyer, 
A.\ G\"orgen, B.\ Hyland, D.\ Jenkins, F.\ Johnston-Theasby, P.\ Joshi, A.\ O.\ Machiavelli,
F.\ Moore, G.\ Mukherjee, A.\ A.\ Phillips, W.\ Reviol, D.\ G.\ Sarantites, 
M.\ A.\ Schumaker, D.\ Seweryniak, M.\ B.\ Smith, J.J.~Valiente-Dob\'on, R.~Wadsworth, 
{\it submitted to Phys. Rev. C}

\bibitem{Kr72-ing} B.\ G.\ Carlsson and I.\ Ragnarsson, American Institute of 
Physics, Conference Proceedings 831, Int. Conference {\it ``Frontiers in Nuclear 
Structure, Astrophysics, and Reactions: FINUSTAR''}, edited by S.\ V.\ Harissopulos, 
P.\ Demetriou, and R.\ Julin, (2006) p.\ 60.

\bibitem{A190} A.\ V.\ Afanasjev, J.\ K{\"o}nig, and P.\ Ring, {\it Phys.\ Rev.} {\bf C60}, 
051303 (1999).
 
\bibitem{CRHB} A.\ V.\ Afanasjev, P.\ Ring, and J.\ K{\"o}nig, {\it Nucl.\ Phys.} {\bf A676}, 
196 (2000).

\bibitem{VRAL.05} D.\ Vretenar, A.\ V.\ Afanasjev, G.\ Lalazissis, and P.\ Ring, 
{\it Phys.\ Rep.} {\bf 409}, 101 (2005).

\bibitem{Giacomo} G.\ de Angelis, C.\ Fahlander, A.\ Gadea, E.\ Farnea, 
W.\ Gelletly, A.\ Aprahamian, D.\ Bazzacco, F.\ Becker, P.\ G.\ Bizzeti, 
A.\ Bizzeti-Sona, F. Brandolini, D.\ de Acu\~{n}a, M.\ De Poli, J.\ Eberth, 
D.\ Foltescu, S.\ M.\ Lenzi, S.\ Lunardi, T.\ Martinez, D.\ R.\ Napoli, 
P.\ Pavan, C.\ M.\ Petrache, C.\ Rossi Alvarez, D.\ Rudolph, B.\ Rubio, 
W.\ Satu{\l}a, S.\ Skoda, P.\ Spolaore, H.\ G.\ Thomas, C.\ A.\ Ur, and R.\ Wyss,
{\it Phys.\ Lett.} {\bf B415}, 217 (1997).


\bibitem{Beng85} T.\ Bengtsson and I.\ Ragnarsson, {\it Nucl.\ Phys.} {\bf A436}, 14 (1985).

\bibitem{A110} A.\ V.\ Afanasjev and I.\ Ragnarsson, {\it Nucl.\ Phys.}  {\bf A591}, 387 (1995).

\bibitem{PhysRep} A.\ V.\ Afanasjev, D.\ B.\ Fossan, G.\ J.\ Lane and I.\ Ragnarsson, {\it Phys.\ 
Rep.} {\bf 322}, 1 (1999).

\bibitem{KR.89} W.\ Koepf and P.\ Ring, {\it Nucl.\ Phys.} {\bf A493}, 61 (1989).

\bibitem{KR.93} J.\ K{\"o}nig and P.\ Ring, {\it Phys.\ Rev.\ Lett.} {\bf 71}, 3079 (1993).

\bibitem{A150} A.\ V.\ Afanasjev, J.\ K\"onig and P.\ Ring, {\it Nucl.\ Phys.} {\bf A608}, 
107 (1996).

\bibitem{NL3} G.\ A.\ Lalazissis, J.\ K\"onig and P.\ Ring, {\it Phys.\ Rev.} {\bf C55}, 
540 (1997).

\bibitem{D1S} J.\ F.\ Berger, M.\ Girod, and D.\ Gogny, {\it Comp.\ Phys.\ Comm.} {\bf 63}, 
365 (1991).

\bibitem{Br70}
D.\ G.\ Jenkins, N.\ S.\ Kelsall, C.\ J.\ Lister, D.\ P.\ Balamuth, M.\ P.\ Carpenter, 
T.\ A.\ Sienko, S.\ M.\ Fischer, R.\ M.\ Clark, P.\ Fallon, A.\ G\"orgen, A.\ O.\ Macchiavelli, 
C.\ E.\ Svensson, R.\ Wadsworth, W.\ Reviol, D.\ G.\ Sarantites, G.\ C.\ Ball, 
J.\ Rikovska Stone, O.\ Juillet, P.\ van Isacker, A.\ V.\ Afanasjev and S.\ Frauendorf, 
{\it Phys.\ Rev.} {\bf C65}, 064307 (2002).  

\bibitem{TWH.98} J.\ Terasaki, R.\ Wyss, and P.-H.\ Heenen, {\it Phys.\ Lett.} {\bf B437}, 
1 (1998).

\bibitem{A250} A.\ V.\ Afanasjev, T.\ L.\ Khoo, S.\ Frauendorf, G.\ A.\ Lalazissis, 
and I.\ Ahmad, {\it Phys.\ Rev.} {\bf C67}, 024309 (2003).

\bibitem{GBI.86} D.\ Galeriu, D.\ Bucurescu, and M.\ Iva\c{s}ku, {\it J.\ Phys. G} {\bf 12}, 329 
(1986).

\bibitem{Fisher} S.\ M.\ Fischer, C.\ J.\ Lister, D.\ P.\ Balamuth, R.\ Bauer, 
J.\ A.\ Becker, L.\ A.\ Bernstein, M.\ P.\ Carpenter, J.\ Durell, N.\ Fotiades, 
S.\ J.\ Freeman, P.\ E.\ Garrett, P.\ A.\ Hausladen, R.\ V.\ F.\ Janssens, D.\ Jenkins, 
M.\ Leddy, J.\ Ressler, J.\ Schwartz, D.\ Svelnys, D.\ G.\ Sarantites, D.\ Seweryniak, 
B.\ J.\ Varley, and R.\ Wyss, {\it Phys.\ Rev.\ Lett.} {\bf 87}, 132501 (2001).

\bibitem{A60} A.\ V.\ Afanasjev, I.\ Ragnarsson and  P.\ Ring, {\it Phys.\ Rev.} 
{\bf C59}, 3166 (1999).

\bibitem{Kr73} N.\ S.\ Kelsall, S.\ M.\ Fischer, D.\ P.\ Balamuth, G.\ C.\ Ball, 
M.\ P.\ Carpenter, R.\ M.\ Clark, J.\ Durell, P.\ Fallon,  S.\ J.\ Freeman, 
P.\ A.\ Hausladen, R.\ V.\ F.\ Janssens, D.\ G.\ Jenkins, M.\ J.\ Leddy, 
C.\ J.\ Lister, A.\ O.\ Macchiavelli, D.\ G.\ Sarantites, D.\ C.\ Schmidt, 
D.\ Seweryniak, C.\ E.\ Svensson, B.\ J.\ Varley, S.\ Vincent, R.\ Wadsworth,
A.\ N.\ Wilson, A.\ V.\ Afanasjev, S.\ Frauendorf, I.\ Ragnarsson and 
R.\ Wyss, {\it Phys.\ Rev.} {\bf C65}, 044331 (2002).


\bibitem{Cu59} C.\ Andreoiu, D.\ Rudolph, C.\ E.\ Svensson, A.\ V.\ Afanasjev, 
J.\ Dobaczewski, I.\ Ragnarsson, C.\ Baktash, J.\ Eberth, C.\ Fahlander, D.\ S.\ Haslip, 
D.\ R.\ LaFosse, S.\ D.\ Paul, D.\ G.\ Sarantites, H.\ G.\ Thomas, J.\ C.\ Waddington, 
W.\ Weintraub, J.\ N. Wilsson and C.-H.\ Yu, {\it Phys.\ Rev.} {\bf C62}, 051301(R) 
(2000). 

\bibitem{Sr76} 
P.\ J.\ Davies, A.\ V.\ Afanasjev, R.\ Wadsworth, C.\ Andreoiu, R.\ A.\ E.\ Austin, 
M.\ P.\ Carpenter, D.\ Dashdorj, S.\ J.\ Freeman, P.\ E.\ Garrett, A.\ G\"orgen, 
J.\ Greene, D.\ G.\ Jenkins, F.\ L.\ Johnston-Theasby, P.\ Joshi, A.\ O.\ Macchiavelli, 
F.\ Moore, G.\ Mukherjee, W.\ Reviol, D.\ Sarantites, D.\ Seweryniak, M.\ B.\ Smith, 
C.\ E.\ Svensson, J.\ J.\ Valiente-Dobon, D.\ Ward, {\it submitted to Phys.\ Rev.\ C}.

\bibitem{Pingst-A30-60} A.\ V.\ Afanasjev, P.\ Ring and I.\ Ragnarsson,
Proc. Int. Workshop PINGST2000 "Selected topics on $N=Z$ nuclei", 
2000, Lund, Sweden, Eds. D.\ Rudolph and M. Hellstr{\"o}m, (2000) 
p.\ 183.

\bibitem{SRR.02} J.\ A.\ Sheikh, P.\ Ring, and R.\ Rossignoli, {\it Phys.\ Rev.} 
{\bf C66}, 044318 (2006).

\bibitem{Kr73Rb74-def} F.\ Johnston-Theasby {\it et al, in preparation, to
be submitted to Phys.\ Rev. C.}

\bibitem{Kr74} J.\ J.\ Valiente-Dob\'on, T.\ Steinhardt, C.\ E.\ Svensson,
I.\ Ragnarsson, A.\ V.\ Afanasjev, C.\ Andreoiu, R.\ A.\ E.\ Austin,
M.\ P.\ Carpenter, D.\ Dashdorj, G.~de~Angelis, F.\ D\"onau, J.\ Eberth, 
E.\ Farnea, P.\ Finlay, S.\ J.\ Freeman, A.\ Gadea, P.\ E.\ Garrett,
A.\ G\"orgen, G.\ F.\ Grinyer, B.\ Hyland, D.\ Jenkins, F.\ Johnston-Theasby,
P.\ Joshi, A.\ Jungclaus, K.\ P.\ Lieb, A.\ O.\ Macchiavelli, F.\ Moore,
G.\ Mukherjee, D.\ R.\ Napoli, A.\ A.\ Phillips, C.\ Plettner,
W.\ Reviol, D.\ Sarantites, H.\ Schnare,  M.\ A.\ Schumaker, R.\ Schwengner,
D.\ Seweryniak, M.\ B.\ Smith, I.\ Stefanescu, O.\ Thelen, R.\ Wadsworth,
D.\ Ward, {\it Phys.\  Rev.\  Lett.} {\bf 95}, 232501 (2005).

\bibitem{Gogny}  M.\ Anguiano, J.\ L.\ Egido and L.\ M.\ Robledo,
{\it Nucl.\ Phys.} {\bf A683}, 227 (2001).

\bibitem{Zn60SD} C.\ E.\ Svensson, D.\ Rudolph, C.\ Baktash, M.\ A.\ Bentley, 
J.\ A.\ Cameron, M.\ P.\ Carpenter, M.\ Devlin, J.\ Eberth, S.\ Flibotte, 
A.\ Galindo-Uribarri, G.\ Hackman, D.\ S.\ Haslip, R.\ V.\ F.\ Janssens, 
D.\ R.\ LaFosse, T.\ J.\ Lampman, I.\ Y.\ Lee, F.\ Lerma, A.\ O.\ Macchiavelli, 
J.\ M.\ Nieminen, S.\ D.\ Paul, D.\ C.\ Radford, P.\ Reiter, L.\ L.\ Riedinger, 
D.\ G.\ Sarantites, B.\ Schaly, D.\ Seweryniak, O.\ Thelen, H.\ G.\ Thomas, 
J.\ C.\ Waddington, D.\ Ward, W.\ Weintraub, J.\ N.\ Wilson, C.\ H.\ Yu, A.\ V.\ 
Afanasjev, and I.\ Ragnarsson, {\it Phys.\ Rev.\ Lett.}, {\bf 82}, 3400 
(1999).

\bibitem{fisch03} S.\ M.\ Fischer, C.\ J.\ Lister and D.\ P.\ Balamuth, 
{\it Phys. Rev.}  {\bf C67}, 064318 (2003).

\bibitem{kel02}  N.\ S.\ Kelsall, C.\ E.\ Svensson, S.\ Fischer,
D.\ E.\ Appelbe, R.\ A.\ E.\ Austin, D.\ P.\ Balamuth, 
G.\ C.\ Ball, J.\ A.\ Cameron, M.\ P.\ Carpenter, R.\ M.\ Clark,
M.\ Cromaz, M.\ A.\ Deleplanque, R.\ M.\ Diamond, J.\ L.\ Durell,
P.\ Fallon, S.\ J.\ Freeman, P.\ A.\ Hausladen, D.\ F.\ Hodgson,
R.\ V.\ F.\ Janssens, D.\ G.\ Jenkins, G.\ J.\ Lane, M.\ J.\ Leddy,
C.\ J.\ Lister, A.\ O.\ Macchiavelli, C.\ D.\ O'Leary, D.\ G.\ Sarantites,
F.\ S.\ Stephens, D.\ C.\ Schmidt, D.\ Seweryniak, B.\ J.\ Varley,
S.\ Vincent, K.\ Vetter, J.\ C.\ Waddington, R.\ Wadsworth,
D.\ Ward, A.\ N.\ Wilson, A.\ V.\ Afanasjev, S.\ Frauendorf,
I.\ Ragnarsson and R.\ Wyss, Proc.\ Int.\ Conf. on ``Frontiers 
of Nuclear Structure'', (Berkeley, California, 2002), AIP 
Conf. Proc. v. 656, Eds. P. Fallon and R. Clark, (Melville, New York, 
2003) p.\ 261.


\bibitem{Ge64-exp} P.\ J.\ Ennis, C.\ J.\ Lister, W.\ Gelletly, H.\ G.\ Price,
B.\ J.\ Varley, P.\ A.\ Butler, T.\ Hoare, S.\ Cwiok, W.\ Nazarewicz,
{\it Nucl.\ Phys.} {\bf A535}, 392 (1991).

















 

\end{thebibliography}
\end{document}